\documentclass[sn-aps,referee]{sn-jnl}
\usepackage{graphicx}

\graphicspath{ {./} }
\usepackage{epsfig} 

\usepackage{listings}%

\lstdefinestyle{mystyle}{
    backgroundcolor=\color{lightgray},   
    commentstyle=\color{green},
    keywordstyle=\color{magenta},
    numberstyle=\tiny\color{gray},
    stringstyle=\color{purple},
    basicstyle=\ttfamily\small,
    breakatwhitespace=false,         
    breaklines=true,                 
    captionpos=b,                    
    keepspaces=true,                 
    numbers=left,                    
    numbersep=5pt,                  
    showspaces=false,                
    showstringspaces=false,
    showtabs=false,                  
    tabsize=2
}

\usepackage{amsmath}
\usepackage{amssymb}
\usepackage{graphicx}
\usepackage{color,soul}
\usepackage{amsthm}
\usepackage{algpseudocode}
\usepackage{algorithm}
\usepackage{verbatim}

\usepackage[belowskip=-15pt,aboveskip=5pt]{caption}
\usepackage{subcaption}
\usepackage{setspace}
\usepackage{graphicx}%
\usepackage{multirow}%
\usepackage{amsmath,amssymb,amsfonts}%
\usepackage{amsthm}%
\usepackage{mathrsfs}%
\usepackage[title]{appendix}%
\usepackage{xcolor}%
\usepackage{textcomp}%
\usepackage{manyfoot}%
\usepackage{booktabs}%
\usepackage{algorithm}%
\usepackage{algorithmicx}%
\usepackage{algpseudocode}%
\usepackage{listings}%
\usepackage{lineno}
\usepackage{titlesec}
\titlespacing*{\section}{0pt}{0.1\baselineskip}{0.2\baselineskip}

\title{Optimal Box Contraction for Solving Linear Systems via Simulated and Quantum Annealing}

\author[1]{\fnm{Sanjay} \sur{Suresh}}\email{ssuresh27@wisc.edu}

\author*[2]{\fnm{Krishnan} \sur{Suresh}}\email{ksuresh@wisc.edu}

\affil[1]{\orgdiv{Computer Science}, \orgname{University of Wisconsin, Madison}, \orgaddress{\street{1210 W. Dayton Street}, \city{Madison}, \postcode{53706}, \state{WI}, \country{USA}}}

\affil*[2]{\orgdiv{Mechanical Engineering}, \orgname{University of Wisconsin, Madison}, \orgaddress{\street{1513 University Avenue}, \city{Madison}, \postcode{53706}, \state{WI}, \country{USA}}}

\begin{document}

\abstract
{ 
Solving linear systems of equations is an important problem in science and engineering. Many quantum algorithms, such as the Harrow-Hassidim-Lloyd (HHL) algorithm (for quantum-gate computers) and the box algorithm (for quantum-annealing machines), have been proposed for solving such systems. 

The focus of this paper is on improving the efficiency of the box algorithm. The basic principle behind this algorithm is to transform the linear system into a series of quadratic unconstrained binary optimization (QUBO) problems, which are then solved on annealing machines.

The computational efficiency of the box algorithm is entirely determined by the number of iterations, which, in turn, depends on the box contraction ratio, typically set to 0.5.  Here, we show through theory that a contraction ratio of 0.5 is sub-optimal and that we can achieve a speed-up with a contraction ratio of  0.2. This is confirmed through numerical experiments where a speed-up between $20 \%$ to $60 \%$ is observed when the optimal contraction ratio is used.
}

\maketitle

\section{Introduction}

Solving least squares problems and linear systems of equations are of utmost importance in science and engineering. Many algorithms have been proposed to solve such problems on classical computers. Quantum computers have recently been proposed as an alternate since they can potentially accelerate the computation \cite{tosti2022review, wang2023opportunities}. In particular, the Harrow-Hassidim-Lloyd (HHL) algorithm is a landmark strategy for solving linear systems of equations on \emph {quantum-gate computers}. In theory, it offers an exponential speed-up over classical algorithms \cite{harrow2009quantum}, and it has been further improved recently \cite{ambainis2010variable,childs2017quantum, liu2022survey, costa2022optimal}. However, due to the accumulation of errors in current noisy intermediate-scale quantum (NISQ) computers  \cite{preskill2018quantum}, the HHL algorithm and its variants are limited, in practice, to tiny systems \cite{ji2020demonstration}. Furthermore, extracting the target state can be expensive/impractical \cite{clader2013preconditioned}.

In parallel, \emph {quantum annealing machines}, such as the D-Wave systems with several thousand qubits \cite{shin2014quantum}, have also been proposed for solving such problems since they are less susceptible to noise \cite{hauke2020perspectives, yarkoni2022quantum}. The basic principle is to convert the least squares and linear system into a series of quadratic unconstrained binary optimization (QUBO) problems. For example, O'Malley and Vesselinov solved the least-squares problem using a finite-precision qubit representation  \cite{o2018nonnegative}.  Borle and Lomonaco conducted a theoretical and numerical analysis of this approach \cite{borle2019analyzing, borle2022viable}. 

In this paper, we consider solving linear system of equations via the QUBO formulation. If the matrix is positive-definite, which is often the case, one can pose this as a energy minimization problem and thereby convert the linear system into a series of QUBO problems. However, current quantum annealing machines are only equipped with about 2000 qubits, with additional restrictions on connectivity \cite{shin2014quantum}. This implies that: (1) the size of the linear system is somewhat limited, and (2) the solution can only be computed with limited precision. The first limitation can be addressed through a hybrid  Gauss-Seidel strategy \cite{pollachini2021hybrid}. In contrast, the second limitation typically relies on the \emph{iterative box algorithm} \cite{srivastava2019box}, the main focus of this paper. Alternate iterative methods are proposed in \cite{raisuddin2022feqa}.

In the box algorithm, the number of iterations strongly depends on the box contraction ratio, i.e., the ratio by which the box size reduces under certain conditions (see Section \ref{sec:overviewQCalgorithms}). This ratio is typically set to 0.5 \cite{srivastava2019box}.  In this paper, we show, through theoretical analysis, that a contraction ratio of 0.5 is sub-optimal and that we can achieve a speed-up with a contraction ratio of (approximately) 0.2. This is confirmed through numerical experiments using simulated and quantum annealing.

\section{Background} 
\label{sec:overviewQCalgorithms}

\subsection{QUBO Formulation}

\begin{figure}[H]
 	\begin{center}
	\includegraphics[width=0.25\textwidth]{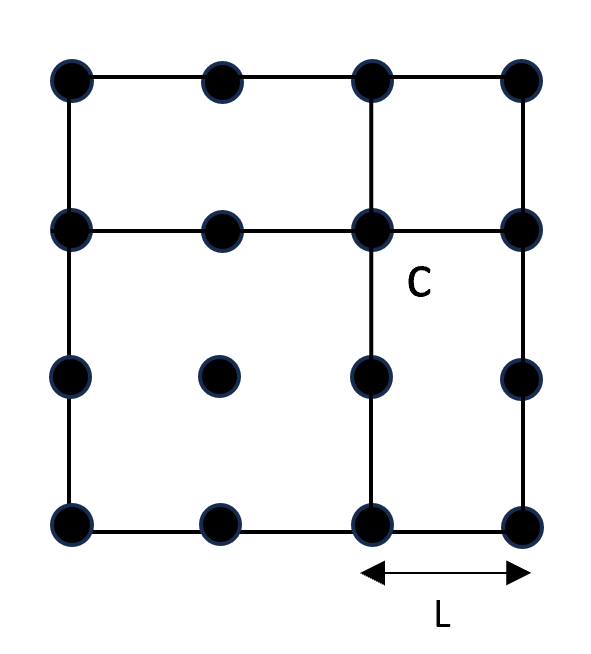}
    \caption{The box representation for $d = 2$.}
	\label{fig:boxRep}
	\end{center}
\end{figure}

Consider the following linear system of equations:

\begin{equation}
\label{reducedProblem} 
\mathbf A \mathbf x = \mathbf b
\end{equation}
where $\mathbf A$ is a $d\times d$ matrix.  If $\mathbf A$ is positive-definite (assumed to be true in the remainder of the paper), then solving Eq. \eqref{reducedProblem} is equivalent to minimizing the potential energy:
\begin{equation}
\label{hamiltonian2} 
    \min_{\mathbf x} \Pi =  \frac 1 2 \mathbf {x}^T\mathbf A \mathbf {x} -  \mathbf {x}^T \mathbf {b} 
\end{equation}
We represent each real component $x_j$ using qubit variables to solve this on a quantum annealing machine. A well-known strategy is the two's complement radix representation \cite{borle2022viable, borle2019analyzing}; for example, a scalar variable $x$ can be represented using $m$ qubits as:
\begin{equation}
    x = -q_12^{m-1} + \sum_{i=2}^{m} q_i 2^{i-2}
\end{equation}
Since the number of qubits is often limited in a quantum machine, it is common to let $m = 2$, leading to:
\begin{equation} 
x = -2q_1+q_2
\end{equation}
Of course, this can only capture the numbers $\{-2,-1,0,1\}$. However, one can easily extend this to a wide range of real numbers through scaling $L$ and offset $c$ via (see Section \ref{sec:BoxAlgorithm} for further explanation):
\begin{equation} 
x = c+L(-2q_1+q_2)
\end{equation}
This is often called the \emph {box representation} \cite{srivastava2019box}. Furthermore, one can easily generalize this to arbitrary dimension $d$ via:
\begin{equation}\label{boxRepresentation}
\mathbf x = \mathbf c + L(-2\mathbf q_1 + \mathbf q_2)
\end{equation}
where $\mathbf q_1 $ and $\mathbf q_2$ are qubit vectors of length $d$, i.e., a total of  $2d$ qubits is used to capture  $ \mathbf {x} $. Thus, a $d$-dimensional system is associated with $4^d$ total states. This is illustrated schematically in Fig. \ref{fig:boxRep} for $d = 2$. 

Since $\mathbf x $  is linear in $\mathbf q_1 $ and $\mathbf q_2$, substituting Eq. \eqref{boxRepresentation} into Eq. \eqref{hamiltonian2} leads to a quadratic unconstrained \emph {binary} optimization (QUBO) problem:
\begin{equation}  
\label{hamiltonianQUBOInitial} 
    \min_{\mathbf q  = \{ \mathbf q_1 , \mathbf q_2 \}} \Pi = \frac 1 2 \mathbf q^T \mathbf Q' \mathbf q  + \mathbf q^T \mathbf r
\end{equation}
Furthermore, since the qubit variables can only take the values $0$ or $1$, the linear term can be absorbed into  the quadratic term \cite{zaman2021pyqubo}, resulting in the standard form: 
\begin{equation}
\label{hamiltonianQUBO} 
    \min_{\mathbf q = \{ \mathbf q_1 , \mathbf q_2 \} } \Pi = \frac 1 2 \mathbf q^T\mathbf Q \mathbf q 
\end{equation}
where $ \mathbf Q$ is symmetric.  

\subsection{Box Algorithm} 
\label{sec:BoxAlgorithm}
\begin{figure}[H]
        \begin{center}
 \begin{subfigure}[c]{0.36\textwidth}	
	\includegraphics[width=1\textwidth]{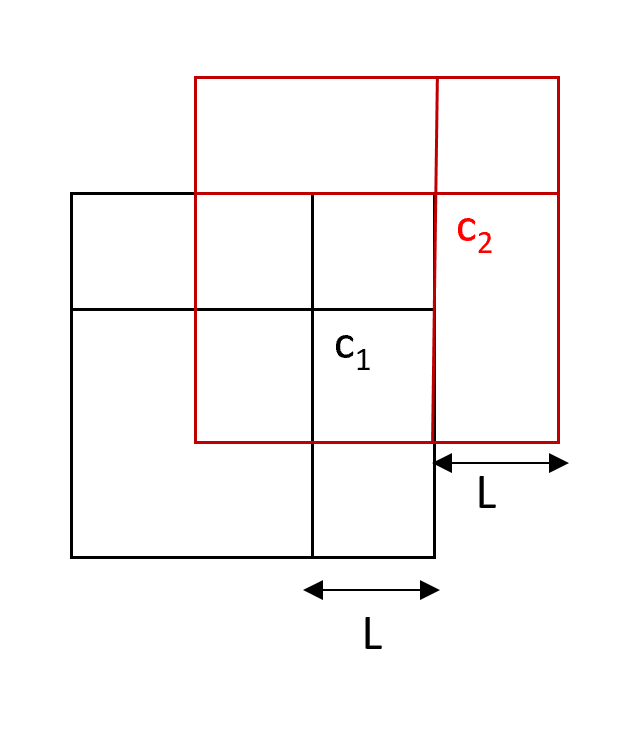}
        \label{fig:boxAlgTranslate}
 \end{subfigure}
     \begin{subfigure}[c]{0.36\textwidth}	
	   \includegraphics[width=0.8\textwidth]{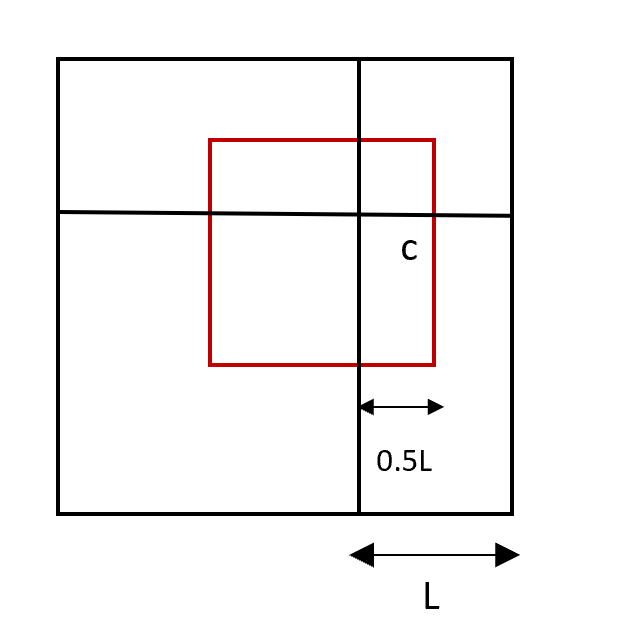}
 \end{subfigure}
 	   \end{center}
\caption{Box algorithm: (a) translation, (b) contraction. }
\label{fig:boxAlg}
 \end{figure}

We now describe the box algorithm (see Algorithm \ref{alg:Box}) that exploits the QUBO formulation to solve Eq. \eqref{reducedProblem} to a high degree of precision.  During each iteration of the algorithm, the center $\mathbf c$ or scale $L$ is updated as follows. In a particular iteration, when a QUBO problem in Eq. \ref{hamiltonianQUBO} is solved, if a lower potential energy state than the current state is reached, $\mathbf{c}$ is updated (referred to as a \emph{translation}; see Fig \ref{fig:boxAlg}a), else  $L$ is reduced by a factor of, typically, $0.5$ (referred to as a \emph{contraction}; see Fig \ref{fig:boxAlg}b). The iteration is then continued until $L$ falls below the precision desired. 

We now describe the box algorithm (see Algorithm \ref{alg:Box}) in detail since it will be relevant for the remainder of the paper. Observe in the algorithm that: 
\begin{itemize}
    \item Lines 2-7: Various quantities are initialized: (1) the center $\mathbf c $  is initialized to $\mathbf 0$, (2) the length $L$ is set to $1$ (see remark below), (3) the qubit vectors $\mathbf q_1 $ and $\mathbf q_2$ are created, (4) the number of translations $N_t$, and contractions $N_t$,  are initialized to $0$, (5) the contraction ratio $\beta$ is initialized to $0.5$ and (5) the potential energy is initialized to 0.
    \item Line 9: The unknown vector $ \mathbf x$ is represented via the qubits and the current  $\mathbf c $   and $L$ .
    \item Line 10: The potential energy $\Pi$ is formulated.
    \item Line 11: The minimum value of $\Pi$ and the corresponding qubit values are determined either via simulated annealing or quantum annealing. 
    \item Lines 12-15: If a lower energy state is found, the center is translated and $N_t$, the number of translations, is incremented (see remark below).
    \item Lines 16-18:  Else, $N_c$, the number of contractions, is incremented and $L$ is reduced by a factor of $\beta$.
    \item Line 20: The algorithm terminates if $L$ is less than the desired tolerance, or if the number of total iterations $(N_t + N_c)$ is greater than an allowable $N_{\text{allowable}}$
\end{itemize}

\underline{Remark on Line 3}: Although $L$ is initialized to $1$, the box algorithm is robust in that it converges for any reasonable value of $L$ \cite{srivastava2019box}. However, choosing $L$ incorrectly will lead to slower convergence. 

\underline{Remark on Line 12}: We have observed that the box algorithm is more stable and unnecessary translations can be avoided if we add a small buffer by checking if {$\Pi^* < \hat \Pi (1+10^{-8} $}). 

\begin{algorithm}[H]
    \caption{Box Algorithm}
    \label{alg:Box}
    \begin{algorithmic}[1]
        \Procedure {BoxAlg}{$\mathbf A$, $\mathbf b$, $\epsilon$,$N_{allowable}$ }
        \State  $ \mathbf c \leftarrow \mathbf 0  $ \Comment{Center of length $d$} \label{alg:center_init}
        \State  $L \leftarrow 1$ \Comment{Initialize box size} \label{alg:length_init}
        \State $\mathbf q_1,\mathbf  q_2   \leftarrow \text{Qubits}(d)   $  \Comment{Create qubit arrays of length d}
        \State $ N_c = N_t = 0$ \Comment{Translation and contraction steps set to 0}        
        \State $\beta = 0.5$ \Comment{Contraction ratio}
        \State $\hat \Pi = 0$ \Comment{Initial potential energy}
        \Repeat \Comment{Until convergence}
        \State $\mathbf {x} \leftarrow \mathbf c + L(-2\mathbf q_1 +\mathbf q_2)$\Comment{Symbolic expression}
        \State $\Pi \leftarrow \frac 1 2 \mathbf {x}^T\mathbf A \mathbf {x} -  \mathbf {x}^T \mathbf {b} $ \Comment{Construct QUBO}
        \State $\Pi^*, \mathbf q_1^*, \mathbf q_2^*  \leftarrow \text{minimize}(\Pi)$ \Comment{Solve QUBO}
        \If {$\Pi^* < \hat \Pi $} \Comment{We have found a lower energy state}
        \State $ \mathbf c \leftarrow \mathbf c + L(-2\mathbf q_1^* +\mathbf q_2^*) $ \Comment{Translation of box}
        \State $N_t = N_t+1$ \Comment{Update translation counter}
        \State $\hat \Pi = \Pi^*$ \Comment{Update the lowest potential energy}
        \Else 
        \State $ L \leftarrow \beta L$ \Comment{Reduce box size} 
        \State $N_c = N_c+1$ \Comment{Update contraction counter}
        \EndIf
        \Until ({$L< \epsilon   $}) or ({$N_c+N_t > N_{allowable}$}) \Comment{Termination}
        \EndProcedure    
        \Comment{Output: solution $\mathbf c$}
    \end{algorithmic}
\end{algorithm}

\begin{figure}[H]
 	\begin{center}
	\includegraphics[width=0.5\textwidth]{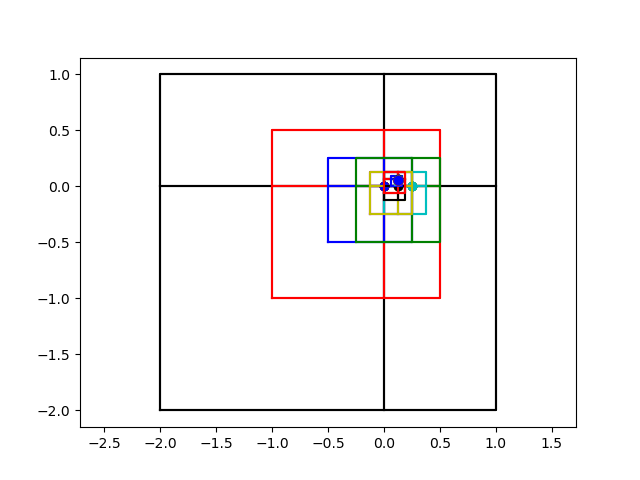}
    \caption{Typical box convergence in 2D.}
	\label{fig:boxConvergence}
	\end{center}
\end{figure}

A typical convergence of the box algorithm in 2D is illustrated in Fig \ref{fig:boxConvergence}. Observe that the number of QUBO problems one must solve is equal to the total number of iterations $(N = N_t + N_c)$. The objective of this paper is to reduce $N$ by finding an optimal value for $\beta$. In particular, we show, in the next section, that the default value of $\beta = 0.5$ recommended in the literature is sub-optimal. 

\section {Box Algorithm Analysis}

\subsection{Contraction steps}

Observe that each time the box contracts, the length of the box reduces by a factor of $\beta$, until $L< \epsilon $. Consequently, the total number of contractions, independent of the number of translations and the linear system being solved, is given by

\begin{equation}\label{contrFormula}
    N_c \sim \log_{\beta} \epsilon
\end{equation}

\subsection{Translation steps}
Although the number of contraction steps is independent of $\beta$, the number of translations depends on $\beta$. Our objective is to determine an upper bound $\hat N_t$ and average estimate $\overline N_t$ in terms of $\beta$ and $\epsilon $.

\subsubsection{1D Box Algorithm}
 \label{sec:1Danalysis}
We will first consider one-dimensional problems that requires only $2$ qubits. Let $L_i$ be the length of the box \emph{before} the $i$-th contraction, i.e., $L_1=1$, $L_2=\beta$, $L_3=\beta^2$, etc. Further, at the start of the algorithm, $c=0$, and $c$ gets updated as follows:

\begin{equation}\label{oneDimQubit}
    c \leftarrow c + L_i(-2q_1 + q_2)
\end{equation}
Let $n_i$ be the number of translation steps \emph{before} the $i^{th}$ contraction. Our objective is to find an upper bound for $n_i$. 

We will assume that the solution $x$ lies within the range $(-2,1)$, else arriving at an upper bound for $N_t$ is impossible. Consider the possible sequence of translations, starting at $c = 0$, before the first contraction. If there is no translation, then $n_1 = 0$, else, $c$ gets updated to $\{-2,-1,1\}$, as per Eq \eqref{oneDimQubit}. Suppose $c\leftarrow -1$, corresponding to $q_1 = 1$, $q_2 = 1$, then the solution $x$ must lie in the range $(-1.5,-0.5)$. Similarly, if $c\leftarrow -2$, or $c\leftarrow 1$, $x$ must lie in the range $(-2,-1.5)$ or $(0.5,1)$ respectively. 

No further translation is possible since it would require $c$ to translate to an inferior solution. In other words, the box must contract in the next iteration. In summary, $n_1\leq 1$.

After this, the box will contract, and $L_2=\beta$, and $x$ must lie in the range $[c-1/2, c+1/2]$, where $c$ is the updated center. Therefore, the next sequence of translations can move the center by at most $0.5$ units where each translation is at least $L_2 = \beta$, as per Eq. \eqref{oneDimQubit}. Note that the center can also translate by $-2 \beta$ per Eq. \eqref{oneDimQubit}, However, since we are seeking an upper bound for $N_t$, we consider the worst-case scenario. Therefore, the maximum number of translations is given by  $n_2 \leq \frac {0.5} {\beta}$. After this, the box must contract, resulting in $L_3 = \beta^2$.

After the contraction, the solution $x$ must lie in the range $[c-\frac \beta {2}, c+ \frac \beta {2}]$, where $c$ is the updated center. The next sequence of translations can move the center by at most $\beta/2$ units and each translation is at least $L_3 =  \beta^2$, as per Eq. \eqref{oneDimQubit}. Consequently, the  maximum number of translations $n_3 \leq (\beta/2)/\beta^2 = \frac {0.5} {\beta}$. After this, the box contracts to $L_4 = \beta^3$. 

\begin{figure}[H]
\begin{center}
\includegraphics[width=0.7\textwidth]{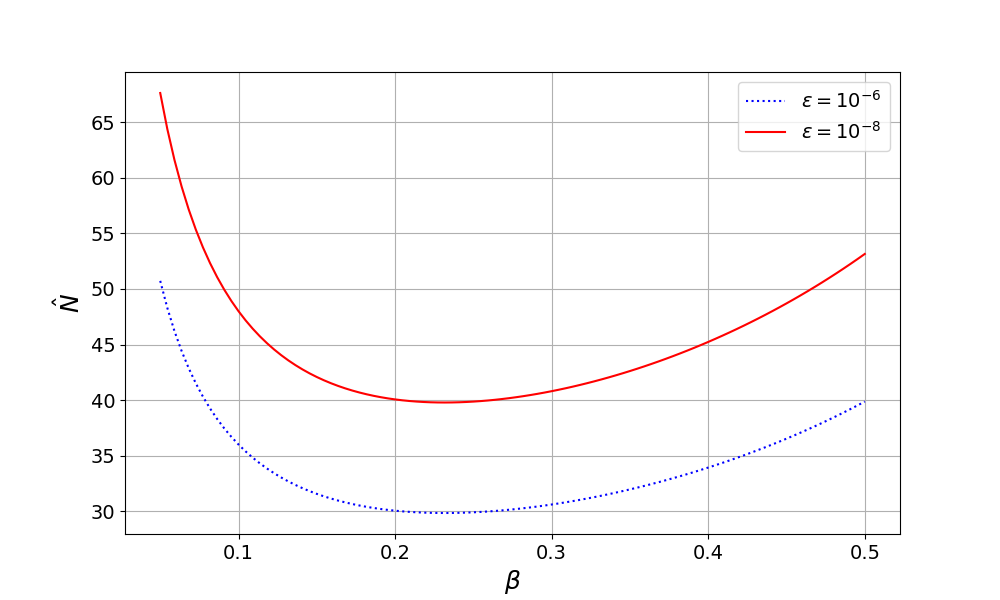}   
\end{center}
\caption{Upper bound $\hat N$ on the number of box iterations.  }
\label{fig:upperBound}
\end{figure}

Repeating this logic for all $N_c$ contractions, we have $n_1\leq 1$ and $n_2,n_3,\dots , n_{N_c} \leq \frac 1 {2\beta}$. Therefore, the total number of translation steps has the following upper bound:

\begin{equation}\label{oneSum}
    \hat{N}_t = \sum_{i=1}^{N_c} n_i = 1 + \frac 1 {2\beta} + \frac 1 {2\beta} + \cdots + \frac 1 {2\beta} = 1+\frac{N_c-1}{2\beta} 
\end{equation}
If we assume $\beta \leq 0.5$ (as we will see below, this is justifiable):
\begin{equation}\label{twoSum}
    \hat{N}_t = \frac{N_c}{2\beta} 
\end{equation}
Combining this with Eq. \eqref{contrFormula}, we have an upper bound on the total iterations
\begin{equation}
    \label{Eq:NHat}
    \hat{N} = \hat{N}_t + N_c = \left( 1+\frac 1 {2\beta} \right)\log_{\beta} \epsilon
\end{equation} 
To find the optimal value of $\beta$, we can the derivative of $\hat N$ in Eq. \ref{Eq:NHat} with respect to $\beta$ and set this equal to $0$, resulting in:
\begin{equation}
    \ln \beta + 2\beta + 1 = 0
\end{equation}
Solving this numerically, we obtain $\beta^* \sim 0.232$, independent of $\epsilon$. This is illustrated in Fig. \ref{fig:upperBound}. For this optimal value, one can observe a $32\%$ reduction in the maximum box iterations, compared to the default value of $\beta = 0.5$. Further, observe that the number of translations increases for $\beta > 0.5$, justifying our earlier assumption.

\begin{figure}[H]
\begin{center}	
\includegraphics[width=0.7\textwidth]{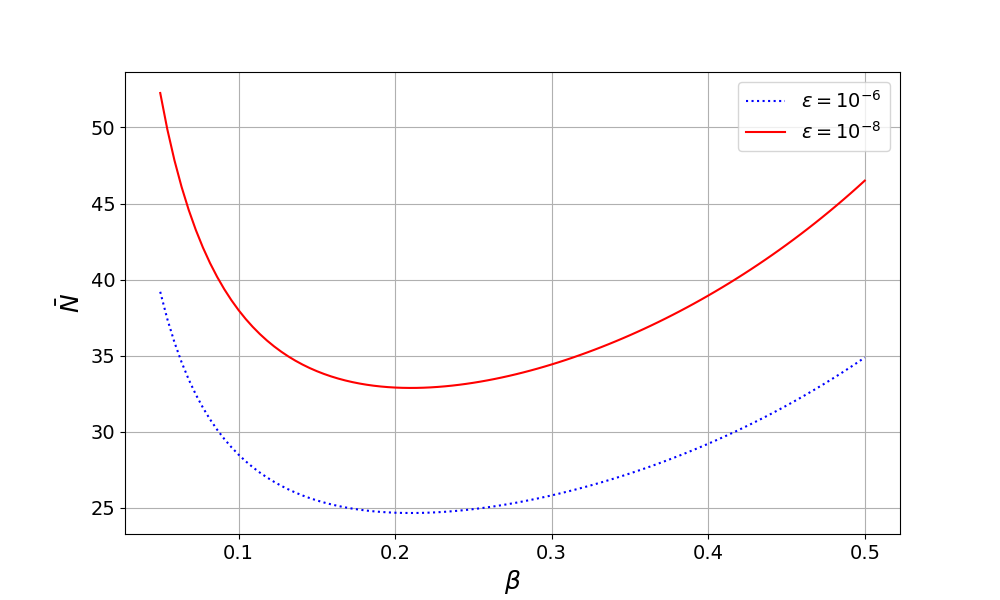}
\end{center}
\caption{Average number of box iterations $\overline N$.  }
\label{fig:AverageIterations}
\end{figure}

Instead of the upper bound, one can also consider the average number of translations $\overline N_t$. There are three different scenarios during each translation ($-1$, $-2$, $1$). If we assume there is an equal probability of translating in each of these directions (a very simplistic model), then the expected translation is given by $( \lvert-1\lvert + \lvert-2\lvert+ \lvert+1\lvert)/3 = 4/3$ (as opposed to $1$ in the worse case). Consequently 
\begin{equation}
     \overline{N}_t =\frac{3N_c}{8\beta}
\end{equation}
Since the number of contractions remains the same, we have:
\begin{equation}
    \overline{N} = \overline N_t + N_c \approx \left( 1+\frac 3 {8\beta} \right)\log_{\beta} \epsilon
\end{equation}
Fig. \ref{fig:AverageIterations}  illustrates $\overline{N}$ vs $\beta$. Taking the derivative of $\overline{N}$ with respect to $\beta$ and setting it equal to $0$, we get $\beta^* \approx 0.21$.  For this optimal value, one can observe a $44\%$ reduction in the average box iterations, compared to the default value of $\beta = 0.5$. 

\subsubsection{Multi-Dimensional Problems}
In this section, we argue that the previous results also holds true in higher dimensions. For a $d$-dimensional problem, let $n_{i,k}$ represent the number of translations in the $k^{th}$ dimension before the $i^{th}$ contraction. Since the dimensions are independent, $n_{i,k}$ are all independent of each other. 

In order to prove this, consider the following hypothetical scenario: Before the $i^{th}$ contraction, let the center translate in the first dimension until it cannot translate anymore in this direction. Then let it translate in dimension $2$, and so on. Once dimension $d$ is reached, this process is repeated until the box no longer can translate in any dimension. Let $n_{i,k}$ be the total number of times the box contracted in each dimension. By the premises of the box-algorithm, the box must now contract. 

Following the logic from the previous section, we have $ \forall i\geq 2, \forall k, n_{i,k} \leq \frac 1 {2\beta}$. However, the maximum number of translations is dictated by one or more of the dimensions. Therefore, let $n_i =  \max_k n_{i,k} $. As a result, the upper bound on translation  is given by
\begin{equation}\label{multiSum}
\hat N_t = \sum_{i=1}^{N_c-1} n_i = \frac{N_c}{2\beta}
\end{equation}
Consequently, 
\begin{equation}\label{multiDimEquation}
    \hat N =   \left( 1+\frac 1 {2\beta} \right)  \log_{\beta}  \epsilon
\end{equation}
irrespective of the number of dimensions. This is later confirmed in the next section through numerical experiments. Similar arguments can be made for the average case.

\section{Numerical Experiments}
\label{sec:Experiments}
We will now carry out numerical experiments to validate the theoretical analysis. The experiments rely heavily on simulated annealing (SA) since quantum annealing (QA) is expensive today. However, a limited number of QA experiments are also carried out. For SA, we rely on D-Wave's Neal annealer; for hybrid QA, we rely on D-Wave's LeapHybridSampler, and for (pure) QA, we rely on DWaveSampler, with EmbeddingComposite. For all three methods, 20 samples were used. 

The QUBO problems were constructed using the pyQUBO package  \cite{zaman2021pyqubo}. The Python code used in generating the results in this section is available from the github link provided towards the end of the paper. 

\subsection{ Positive Definite Matrices}

\begin{figure}[H]
	\begin{center}
     \begin{subfigure}[c]{0.65\textwidth}	
	   \includegraphics[width=1\textwidth]{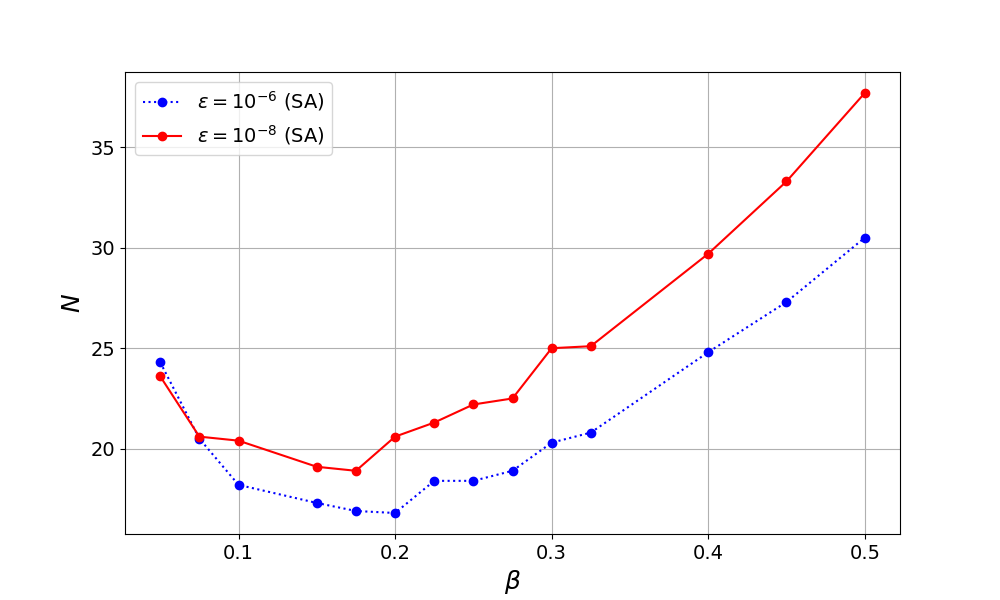} 
        \label{fig:d=2}
     \end{subfigure}
 \end{center}
       \begin{center}
 \begin{subfigure}[c]{0.65\textwidth}	
	\includegraphics[width=1\textwidth]{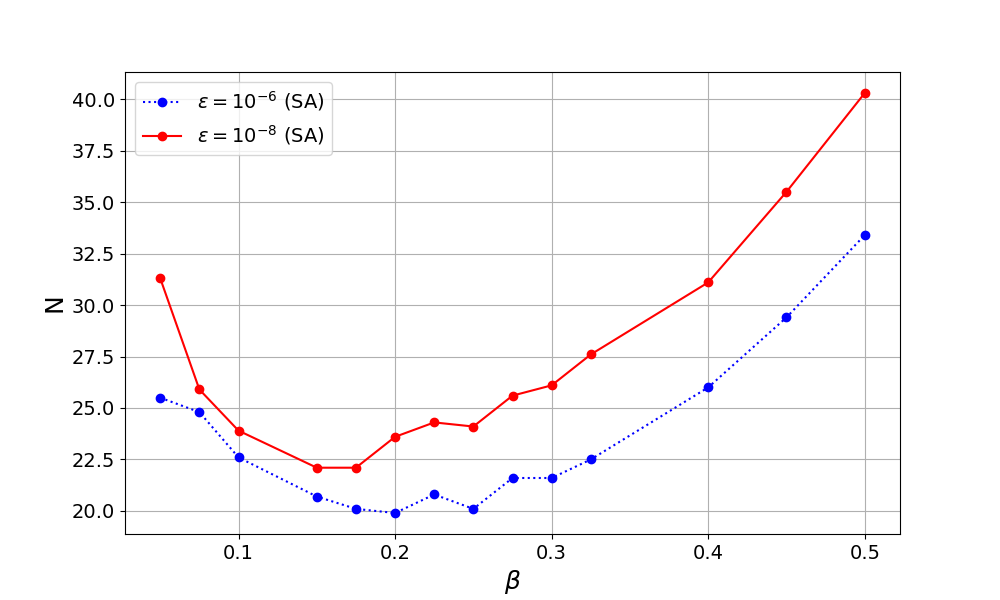}
        \label{fig:d=10}
 \end{subfigure}
         \end{center}
         \begin{center}
  \begin{subfigure}[c]{0.65\textwidth}	
	\includegraphics[width=1\textwidth]{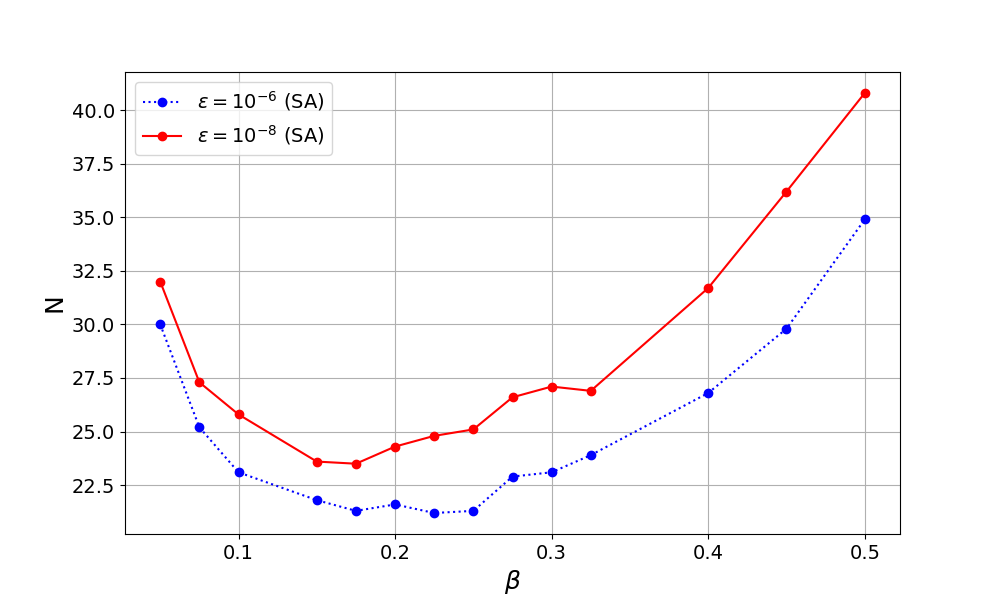}
        \label{fig:d=20}
 \end{subfigure}
   \end{center}
\caption{Observed $N$ vs $\beta$ averaged over ten $d \times d$ problems: (a) d = 2, (b) d = 10, and (c) d = 20. }
\label{fig:PDResults}
\end{figure}

In the first set of experiments, we generate random $d$-dimensional positive definite matrices $\mathbf A$. Further, since we require $\mathbf x$ to lie within $[-2,1]^d$, we first generate $\mathbf x$, and then construct the corresponding right-hand-side $\mathbf b$. The corresponding Python code is given below:
\begin{lstlisting}[language=Python, caption= Generating d-dimensional positive definite matrices and right-hand side.]
B = np.random.rand(d, d)
A = d*np.eye(d)-(B + B.transpose())/2
x = np.array([random.uniform(-2, 1) for _ in range(d)])
b = A.dot(xExact)
\end{lstlisting}

In order to capture the average behavior of the box algorithm, we create ten instances of $\mathbf A$ and $ \mathbf b$, for  $ d = 2, d = 10$ and $ d = 20 $. Finally, for each instance, we use the box algorithm (see Algorithm \ref{alg:Box}) to solve for $\mathbf x$ for $\epsilon = 10^{-6}$ and $\epsilon = 10^{-8}$, for various values of $\beta$. All experiments in this section are carried out using SA. The results are summarized in Fig. \ref{fig:PDResults}. Observe the following:
\begin{itemize}
\item All three graphs exhibit a minima around $\beta = 0.2$, independent of the dimension $d$ and desired accuracy $\epsilon$.
\item The number of iterations is (nearly) independent of the dimension of the problem.
\item The number of iterations is closer to the theoretical prediction $\overline N$ in Fig. \ref{fig:AverageIterations} than to the upper bound prediction $\hat N$ in Fig. \ref{fig:upperBound}.
\end{itemize}

\subsection{1D Poisson Problem}
\begin{figure}[H]
 	\begin{center}
	\includegraphics[trim={0 0 0 0},width=0.7\textwidth]{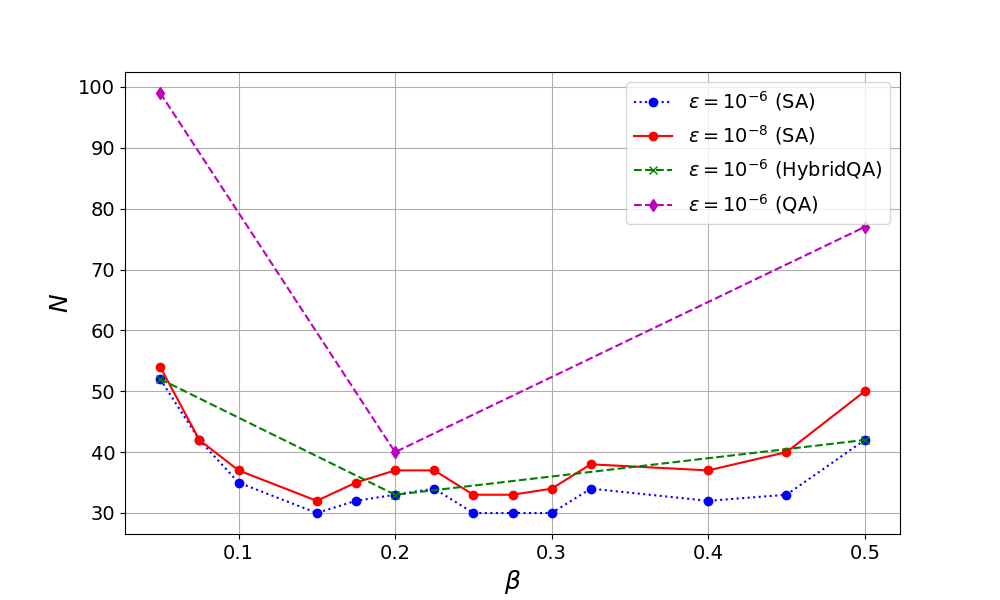}
            \caption{Observed $N$ versus $\beta$ for a 1D Poisson problem. }
	\label{fig:PoissonProblem}
	\end{center}
\end{figure}

For the next experiment, we construct a $ 6 \times 6$ matrix $\mathbf A$ that arises from a finite difference formulation of 1D Poisson problem \cite{conley2023quantum}:
\begin{lstlisting}[language=Python, caption= Generating a 6-dimensional finite difference matrix and right hand side.]
A = np.array([[6,-6,0,0,0,0],[-6,12,-6,0,0,0],
[0,-6,12,-6,0,0],[0,0,-6,12,-6,0],
[0,0,0,-6,12,-6],[0,0,0,0,-6,12]])
xExact = np.array([-np.pi/9, np.pi/11, -np.pi/20,
np.pi/8,  0.05*np.pi, -np.pi/5 ])
b = A.dot(xExact)])
\end{lstlisting}
The results for SA, hybridQA, and QA are summarized in Figure \ref{fig:PoissonProblem}. Observe that:
\begin{itemize}
\item The overall behavior of $N$ vs. $\beta$ is consistent with the theory. 
\item The hybridQA results precisely match that of SA for the three sampled points, suggesting that D-Wave probably relied entirely on CPU for this scenario. 
\item  QA performed poorly compared to SA or hybridQA. This is consistent with the observations in \cite{borle2022viable}. However, even in this case, a $50 \%$ improvement in performance can be observed for $\beta = 0.2$, compared to $\beta = 0.5$.

\end{itemize}

\section{Conclusions}

The box algorithm is a popular method for solving linear systems of equations via the QUBO formulation. In this paper, a theoretical analysis of the box algorithm was carried out that suggested that a computational speed-up can be easily achieved by making a simple modification to the algorithm. Specifically, the theory suggests that a $43 \%$ speed-up can be obtained by reducing the box contraction ratio from 0.5 to 0.2. This was confirmed through numerical experiments where a speed-up between $20 \%$ to $60 \%$ was observed. Additional experiments involving a larger class of linear systems are needed to corroborate these results. 

While the paper focused on linear systems, the strategy can be extended to least squares systems, and other direct methods for solving linear systems via the QUBO formulation \cite{raisuddin2022feqa}.  Further, the analysis here was restricted to the case when only 2 qubits are used to represent each scalar variable. The extension to the more general case needs to be investigated.  Finally, as observed in \cite{borle2022viable}, while it is possible to exploit quantum annealing to solve linear systems of equations, it is currently limited to small systems and low precision.
\section*{Compliance with ethical standards}
\label{sec:ethics}
The authors declare that they have no conflict of interest.

\section*{Replication of Results}
\label{sec:replic}
The Python code pertinent to this paper is available at \href{https://github.com/UW-ERSL/SPAI/}{https://github.com/UW-ERSL/QUBOBoxContraction}.

\section*{Acknowledgments}
We would like to thank the Graduate School of UW-Madison for the Vilas Associate grant.

\bibliography{QCLinearSolver}

\end{document}